\documentclass{elsart}
\usepackage{amssymb}
\usepackage{graphicx,amssymb,lineno}

\begin{document}
\begin{frontmatter}
\title{Specific Density Of Binding Energy Of Core In $\beta$ - Stable Nuclei Is
2.57 $\rm MeV/fm^3$}

\author{G. K. NIE}

\address{Institute of Nuclear Physics, Ulugbek, Tashkent 702132, Uzbekistan}
\ead{galani@Uzsci.net}
\begin{abstract}

Recently an $\alpha$-cluster model based on the pn-pair interactions with using the
isospin invariance of nuclear force has been proposed. According to the model the
excess neutron pairs fill out  the free space in the core determined by the
difference in the charge  and  matter radii of the $\alpha$-clusters. Then the
number of excess neutrons in $\beta$-stable nuclei depends on the number of the core
$\alpha$-clusters. In such a representation the specific density of binding energy
of core $\rho$ is the only parameter to fit the experimental binding energies of
$\beta$-stable nuclei and it turned out to be a constant value equal to 2.57 $\rm
MeV/fm^3$. Knowing the value $\rho$ allows one to estimate the size of a nucleus
from its experimental binding energy.
\end{abstract}
\end{frontmatter}

\section{Binding Energy Of Core}

The idea that the main properties of nuclei can be described from a simple
representation (like liquid drop \cite{1} or some regular forms of $\alpha$-cluster
structure \cite{2}) has been popular since the very beginning of nuclear physics.
The main features of nuclear force, its strength and the short range, allow one to
find some simple formulas to describe size and binding energies of nuclei.

Recently an $\alpha$-cluster model based on pn-pair interactions with using isospin
invariance of nuclear force has been proposed \cite{3,4,5}. In the framework of this
model new formulas to calculate radii and binding energies of $\beta$-stable nuclei
have been found. Also the model provides some reasonable explanation of existing
excess neutrons in stable nuclei. The spinless nn-pairs of excess neutrons fill out
the free space in the core which is determined by the difference in the volumes
occupied by the charge and  by the matter of the $\alpha$-clusters. The proton
charge radius is bigger than the neutron radius due to isospin invariance of nuclear
force \cite{6}.

In such a representation the value of the charge radius $R$ of an even $Z$ nucleus
can be obtained from an estimation of the volumes  occupied by the alpha-clusters of
the core $N_{core\alpha}$ and the peripheral $\alpha$-clusters $N_{p\alpha} =
N_\alpha - N_{core\alpha}$ ($N_{p\alpha} = 1 \div 5$) , $N_\alpha = Z/2$,

\begin{equation}
R^3 = r_\alpha^3 N_{core\alpha}  + r_{^4\rm He}^3 N_{p\alpha}, \label{1}
\end{equation}
where the radius of  a peripheral $\alpha$-cluster  equals the experimental radius
of the nucleus $^4$He   $r_{^4\rm He} = 1.71 $ fm \cite{7}. In case of
$N_{core\alpha} = 0$, $N_\alpha = N_{p\alpha}$, $R = r_{^4\rm He} N_{\alpha}^{1/3}$
[ 3-5 ]. The value of  the radius of an $\alpha$ - cluster of core $r_\alpha = 1.60$
fm is obtained  from fitting the experimental radii of the nuclei with
$N_{core\alpha} >> N_{p\alpha}$ with the formula $R = r_\alpha N_\alpha ^{1/3}$.

For odd $Z_1 = Z + 1$ the number of $\alpha$-clusters $N_{1\alpha}$ in the nucleus
is $N_{1\alpha} = N_\alpha + 0.5$  and the number of peripheral $\alpha$-clusters is
$N_{1p\alpha} = N_{p\alpha} + 0.5$. Then the formula is

\begin{equation}
R_1^3 =  r_\alpha^3 N_{core\alpha} + r_{^4\rm He}^3 N_{1p\alpha}.\label{2}
\end{equation}

Whereas the nuclear radii are calculated by means of estimation of the volumes
occupied by the $\alpha$-clusters of core and by the $\alpha$-clusters of periphery,
the binding energy is calculated on the total amount of $\alpha$-clusters $N_\alpha$
[5] disregarding to the fact that some of them belong to the core and the others are
of the nucleus periphery. In this paper the model [5] is developed to have a
consistency between the ways of how the radii and the binding energies are
calculated.

In the representation with dividing nucleus for core and periphery the binding
energy of a nucleus is to be the sum  of  the binding energy of the core $E_{core}$
and the internal binding energy of the compound peripheral cluster $E_{N_{p\alpha}}$
consisting of $N_{p\alpha}$ $\alpha$-clusters minus the Coulomb energy of  the
compound cluster interaction with the core $\alpha$-clusters $E^C_{{N_{p\alpha}}
N_{core\alpha}} =  2 N_{p\alpha} 2N_{core\alpha} e^2 / R_p$, where  $R_p$ is the
radius of the last alpha-cluster position in the nucleus (in the center of core mass
system)[5]($R_{p1}$ is the radius of the single pn-pair's position in the odd $Z_1 =
Z + 1$ nucleus),
\begin{equation}
R_p = 2.168 (N_\alpha - 4)^{1/3}; R_{p1} = 2.168 (N_\alpha +0.5 - 4)^{1/3}\label{3}
\end{equation}
So the new formula to calculate binding energy is to be as follows
\begin{equation}
E = E_{core}  +  E_{N_{p\alpha}}  - E^C_{N_{p\alpha} N_{core\alpha}}.\label{4}
\end{equation}

The energy of the peripheral compound cluster $E_{N_{p\alpha}}$  consisting of
$N_{p\alpha}$ $\alpha$-clusters is taken  equal to the  experimental binding energy
of the nucleus  $^8$Be ($N_{p\alpha} = 2$) $E_{^8\rm Be}$= 56.5 MeV, $^{12}$C
($N_{p\alpha} = 3$) $E_{^{12}\rm C}$= 92.2 MeV, $^{16}$O ($N_{p\alpha} = 4$)
$E_{^{16}\rm O}$= 127.6 MeV, $^{20}$Ne ($N_{p\alpha} = 5$) $E_{^{20}\rm Ne}$= 160.6
MeV.

In this representation a nucleus   $A (Z, N + \Delta N)$ with  even $Z$, $N=Z$ and
$\Delta N$ is an even number of excess neutrons,   has the same core as the nucleus
$A_1(Z_1, N_1 + \Delta N +1)$ with $Z_1=Z + 1$, $N_1 = Z_1$. Then $A_1 = A + 3$. In
case of the odd $Z_1$ one excess neutron is glued to the single pn-pair, which is
bound with the three nearest peripheral clusters [4, 5]. The long range Coulomb
energy  of the single pn-pair interaction with the core is compensated with its
contribution to the surface tension energy ( see section 2). So for the odd-odd
nuclei the binding energy is calculated as follows

\begin{equation}
E_1 = E_{core}  +  E_{N_{1p\alpha}} -  E^C_{N_{p\alpha} N_{core\alpha}}, \label{4}
\end{equation}
where $E_{N_{1p\alpha}}$ is the experimental binding energy of the nucleus  $^7$Li
$(N_{1p\alpha} =1.5)$ $E_{^7\rm Li}$= 39.2 MeV, $^{11}$B $(N_{1p\alpha} =2.5)$
$E_{^{11}\rm B}$= 76.2 MeV, $^{15}$N $(N_{1p\alpha} =3.5)$ $E_{^{15}\rm N}$= 115.5
MeV, $^{19}$F $(N_{1p\alpha} =4.5)$ $E_{^{19}\rm F}$= 147.8 MeV and $^{23}$ Na
$(N_{1p\alpha} =5.5)$ $E_{^{23}\rm Na}$= 186.6 MeV.

The binding energy of a core occupying the volume $V_{core} = 4/3 \pi r_\alpha^3
N_{core\alpha}$ can be expressed by a formula with using the specific density of
binding energy $\rho $ [$\rm MeV/fm^3$]

\begin{equation}
E_{core}   = V_{core} \rho.\label{5}
\end{equation}

The binding energy of $N_{core\alpha}$  $\alpha$-clusters $E_{N_{core\alpha}}$ is
easily calculated in the framework of the $\alpha$-cluster model (see section 2),
where the energy of short range nuclear force $E^{nuc}$, the energy of surface
tension $E^{ST}$  and the Coulomb energy $E^C$  are calculated on the number of
$\alpha$-clusters. Then

\begin{equation}
E_{\Delta N}   = E_{core}   - E_{N_{core\alpha}}.\label{6}
\end{equation}

The binding energy of excess neutrons $E_{\Delta N}$   depends only on the number of
the excess nn - pairs $N_{nn}$ \cite{5} (see also section 2), which means that only
some particular number of $\Delta N = 2 N_{nn}$ can have place in the core. This
allows one to find the correspondence between $\Delta N$ and $N_{p\alpha}$. It
brings a result that the specific density of core biding energy is a constant value
for all nuclei of the $\beta$-stability valley and its vicinity.

\section{Binding Energy Of Core $\alpha$ - Clusters}

The $\alpha$-cluster model \cite{3,4,5} has been developed  on  the basis of the
fact that the radii of the most abundant isotopes are determined by the number
$N_\alpha$ \cite{8} and that the binding energies of symmetrical even $Z$ nuclei are
calculated by the formula \cite{9}

\begin{equation}
E =  N_\alpha \epsilon_\alpha + 3(N_\alpha - 2) \epsilon_{\alpha \alpha} ,\label{8}
\end{equation}
where $\epsilon_{\alpha}$ = 28.296 MeV is the experimental energy of the nucleus
$^4\rm He$. The value $3 (N_\alpha - 2)$ is considered as the number of short range
bonds between nearby alpha-clusters and $\epsilon_{\alpha \alpha}$ = 2.425 MeV. In
case of an odd  $Z_1= Z + 1$ symmetrical nucleus the single pn-pair is glued to the
three nearby peripheral clusters with 6 bonds with their six pn-pairs. Thus, for the
symmetrical odd nuclei [9]  the binding energy $E_1$ is calculated as follows

\begin{equation}
E_1 = E + \epsilon_{pn} +  6 \epsilon_{pn pn},\label{9}
\end{equation}
where $\epsilon_{pn}$ = 1.659 MeV and $\epsilon_{pn pn}$ =2.037 MeV. What was
remarkable in [9] that in (8) and (9) the energy portions $\epsilon_{\alpha}$,
$\epsilon_{\alpha \alpha}$, $\epsilon_{pn}$ and $\epsilon_{pn pn}$ were obtained
from analysis of the lightest nuclei with $Z\leq 6$. The binding energy of nuclear
force of an $\alpha$-cluster $\epsilon^{nuc}_{\alpha}$ = 29.060 MeV  was found from
the relation $\epsilon_{\alpha} = \epsilon^{nuc}_{\alpha} - \epsilon^{C}_{\alpha}$
where $\epsilon^{C}_{\alpha} = \Delta E_{np} = 0.764 MeV$, $\Delta E_{np}$ is the
difference between the binding energies of the last neutron and the last proton in
the nucleus $^4$He. The energy of nuclear force interaction $\epsilon^{nuc}_{\alpha
\alpha}$ = 4.350 MeV and the Coulomb energy $\epsilon^{C}_{\alpha \alpha}$ = 1.925
MeV in a short range bond between two nearby $\alpha$-clusters were obtained from
the analysis of the experimental binding energies of the lightest nuclei using the
relation $\epsilon_{\alpha \alpha} = \epsilon^{nuc}_{\alpha \alpha} -
\epsilon^{C}_{\alpha \alpha}$ [5].

The Eqs. (8) and (9) mean that the long range Coulomb interactions between
alpha-clusters in the nuclei with $N_\alpha \geq 5$ (for the nuclei with $N_\alpha
\leq 4$ there are only short range interactions) must be  compensated by the surface
tension energy. From this assumption some formulas to calculate the radius of the
last $\alpha$-cluster position in the nucleus (3), the Coulomb radius of the nucleus
$R^C = 1.869 N_\alpha^{1/3}$ and the formula to calculate the Coulomb energy of the
charge sphere of the radius $R^C$ were obtained \cite{5}.

\begin{equation}
E^C  = 1.849 (N_\alpha)^{5/3}  ;  E_1^C  = 1.849 (N_\alpha + 0.5)^{5/3}. \label{10}
\end{equation}

The surface tension energy $E^{ST}$ is the sum of the square radii of the $N_\alpha
-4$ alpha-clusters' positions ( Eq. (9) in \cite{5}).

A phenomenological formula to calculate the binding energy of the excess neutron
pairs $E_{\Delta N}$ was found from fitting the experimental separation energies of
27 nn-pairs, see Eq. (13) in Ref. \cite{5}.

Thus, the formula to calculate binding energy is  \cite{5}

\begin{equation}
E_{th} =  E^{nuc} + E^{ST} - E^C + E_{\Delta N}.\label{11}
\end{equation}

The Eq. (11), obtained from analysis of a reduced amount of nuclei (the symmetrical
nuclei with $Z \leq 22$ for which the experimental values of $\Delta E_{np}$ are
known), turned out to be good for all $\beta$-stable nuclei. The accuracy is a few
MeV, which is the same as that of Weizs\"{a}cker formula, although the ways of
calculations are different.

One of the important conclusions of the model is that the energy of excess neutrons
$E_{\Delta N}$ depends only on the number of excess neutron pairs $N_{nn}  = \Delta
N /2$. To make the calculations easy we propose a simple approximation to the
phenomenological formula (13) in Ref. \cite{5}

\begin{equation}
E_{\Delta N}   =  (21.93 - 0.762 N_{nn}^{2/3}) N_{nn}.\label{12}
\end{equation}

The values of   $E_{\Delta N}$ are given in Fig. 1 in comparison with the values
calculated by the Eq. (13) of Ref. \cite{5}. In the figure some empirical values
obtained from the Eq. $E_{exp} - (E^{nuc} - E^C + E^{ST})$ where the values
($E^{nuc} - E^C + E^{ST}$) are calculated in the framework of the model are given
too. One can see from the figure that the empirical values of $E_{\Delta N}$ depend
only on the number of $N_{nn}$ disregarding to $Z$.
\begin{figure}
\begin{center}
\includegraphics*[width=10cm]{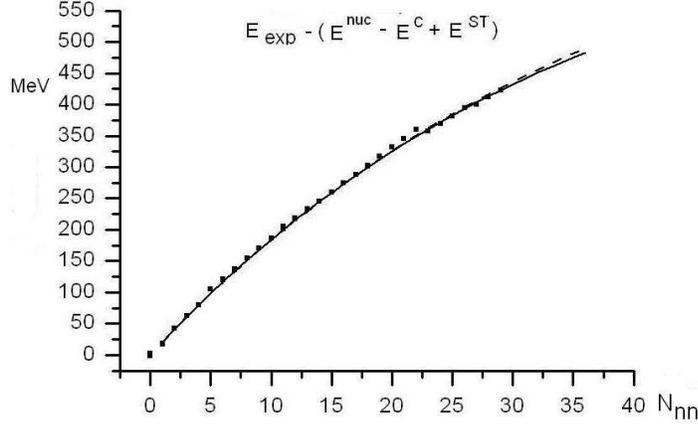}
\end{center}
\caption{The values $E_{exp} - (E^{nuc} - E^C + E^{ST})$ for all $\beta$-stable
even-even nuclei with $Z$ = 10, 20, 30, 40, 50, 60, 70, 80, 90, 100 (41 square
points), where $E^{nuc} - E^C + E^{ST}$ is calculated on the $\alpha$-cluster model
\cite{5}. The solid line indicates the values $E_{\Delta N}$ calculated on the Eq.
(13) in \cite{5}. The dashed line denotes the values $E_{\Delta N}$ (12) of this
paper.}\label{fig:1}
\end{figure}

The binding energy  $E_{N_{core\alpha}}$ of core alpha -clusters  $N_{core\alpha}$
must include the energy of nuclear force of the core $\alpha$-clusters
$E^{nuc}_{N_{core\alpha}}$, the Coulomb energy $E^C_{N_{core\alpha}}$ and surface
tension energy $E^{ST}$. The energy of nuclear force of the core $\alpha$-clusters
includes $N_{core\alpha}\epsilon^{nuc} + 3(N_{core\alpha} - 2)\epsilon^{nuc}_{\alpha
\alpha}$. Also the energy of the number of bonds $\Delta $ between the core
alpha-clusters and the peripheral compound cluster consisting of $N_{p\alpha}$
alpha-clusters must be taken into account

\begin{equation}
\Delta  =  3 (N_\alpha - 2) - 3 (N_{core\alpha} - 2) - 3 (N_{p\alpha} -
2).\label{13}
\end{equation}

One can see that for $N_{p\alpha} \geq 2$  $\Delta_{bonds}  = 6$. So the nuclear
force energy $E^{nuc}_{N_{core\alpha}}$ is calculated as follows

\begin{equation}
E^{nuc}_{N_{core\alpha}} = N_{core\alpha} \epsilon^{nuc} + 3 (N_{core\alpha} - 2)
\epsilon^{nuc}_{\alpha \alpha}+ \Delta \epsilon^{nuc}_{\alpha \alpha}.\label{14}
\end{equation}

The Coulomb energy of core  is equal to the Coulomb energy of $N_{core\alpha}$
$\alpha$-clusters, which is calculated as $1.849 N_{core\alpha}^{5/3}$ due to (10)
plus the Coulomb energy of the $\Delta $  bonds  (13)

\begin{equation}
E^C_{N_{core\alpha}} = 1.849 N_{core\alpha}^{5/3}   + \Delta \epsilon^C_{\alpha
\alpha}.\label{15}
\end{equation}

Using the  relation $\epsilon_{\alpha \alpha}  = \epsilon^{nuc}_{\alpha \alpha} -
\epsilon^C_{\alpha \alpha}$ the formula to calculate the binding energy of the
$N_{core\alpha}$ alpha - clusters is as follows

\begin{equation}
E_{N_{core\alpha}} =   N_{core\alpha}\epsilon^{nuc}_{\alpha} + 3 (N_{core\alpha} -
2)\epsilon^{nuc}_{\alpha \alpha} - 1.849 N_{core\alpha}^{5/3} + \Delta
\epsilon_{\alpha \alpha} + E^{ST}. \label{16}
\end{equation}

We use here the simple function proposed in  \cite{5},  which provides a good
approximation to the phenomenological formula of calculation of the surface tension
energy $E^{ST}$

\begin{equation}
E^{ST}= (N_\alpha + 1.7 ) (N_\alpha - 4)^{2/3} ;  E_1^{ST}= E^{ST} + 1.1(N_\alpha -
4)^{2/3}. \label{17}
\end{equation}

The value of  $1.1(N_\alpha - 4)^{2/3}$ is the contribution of the single pn-pair
into the surface tension energy. But the energy of the Coulomb interaction of the
pn-pair with the $\alpha$-clusters of the core $2N_{core\alpha} e^2 / R_{p1}$ almost
compensates it. For $N_\alpha < 59$ the difference is within few MeV. Therefore to
calculate the binding energy of the odd-odd nuclei in Eq. (5) the Coulomb energy
$E^C_{N_{p\alpha} {N_{core\alpha}}}$ is used.

To fit experimental data for heavy nuclei one has to use a representation that the
peripheral  compound cluster consists of two compound clusters of smaller size. In
the formulas the values of $\Delta $ is changed. For example, for two peripheral
compound clusters $N_{p\alpha 1}$ = 2  and  $N_{p\alpha 2}$ = 3 with the total
amount of alpha-clusters in them $N_{p\alpha} =  5$ the total number of bonds is 1 +
3 = 4. In (13) the number 3 (5 - 2) = 9 is changed for 4, which leads to $\Delta $ =
6 + 5 =11. In (4) $E_{N_{p\alpha}} = E_{N_{\alpha p1}} + E_{N_{\alpha p2}}$ where
$E_{N_{\alpha p1}} = E_{^8\rm Be}$ and $E_{N_{\alpha p2}} = E_{^{12}\rm C}$.

In calculation of the energy $E^C_{N_{p\alpha} N_{core\alpha}}$ one has to take into
account the long range interactions between $N_{\alpha p1}$ and  $N_{\alpha p2}$
clusters. The algorithm looks simple if  one uses a two dimensional matrix
$E^C(N_\alpha, N_{p\alpha})$ for $N_\alpha = 9 \div 59$ and $N_{\alpha p} = 1 \div
8$, which is

\begin{equation}
E^C(N_\alpha, N_{p\alpha}) =  2 N_{p\alpha} (2N_{\alpha}  - 2 N_{p\alpha}) e^2 /
(2.168 ((N_\alpha - 4)^{1/3}). \label{18}
\end{equation}

Then the Coulomb energy for the case is easily expressed as

\begin{equation}
E^C_{N_{p\alpha} N_{core\alpha}} = E^C(N_\alpha, 2) + E^C(N_{\alpha} - 2, 3).
\label{19}
\end{equation}

So, the binding energy of the nucleus $A (Z, N + \Delta N)$  is calculated by the
following way. For $\Delta N$  the value $N_{p\alpha}$ is calculated from Eq. (7)
and Eq. (12) taking into account that the number of nn-pairs in core $N_{nn}$ is
integer. Then the binding energy is calculated by (4) and (5). The radii are
calculated by (1) and (2).  In this approach the expected accuracy is a few MeV,
because the single particle effects (shell effects) are not taken into account. The
value $\rho = 2.57\rm MeV' /fm^3$ fits the binding energies of all $\beta$-stable
isotopes and not stable isotopes of the vicinity. In Table 1 the results of
calculation by (4) and (5) are given for some nuclei. The values of $E_{th}$ (11)
are also presented.

If the value of the specific density of binding energy is known, it gives an
opportunity to estimate the size of the isotopes from their experimental binding
energies. Then the value $N_{p\alpha}$ is calculated by (4) where instead of $E$ the
value $E_{exp}$ is used. For example, the nucleus $^{142}\rm Nd_{60}$ has $E_{exp}$
= 1185 MeV. The value $N_{p\alpha} = 2 \div 5$ are tried and only $N_{p\alpha}$ = 4
corresponds to the $\rho = 2.57$ $\rm MeV /fm^3$. So the radius (1) $R$ = 5.02 fm
and for $^{145}\rm Pm_{61}$ (2) $R$ = 5.05 fm.

\begin{center}{Table 1. Binding energies and radii calculated for $\beta$-stable isotopes.
For stable nuclei the most abundant isotopes have been selected. Also the results
are presented for the corresponding nucleus having the same core (if it is a $\beta$
- unstable nucleus, it is marked by *)}.
\end{center} {\begin{tabular}{ccccccccc}\hline
$Z$&$\Delta N$&$A$&$N_{p\alpha}$&$E_{exp}$\cite{10}&$E$(4,5)&$E_{th}$(11)&$R_{exp}$\cite{11,12,13}&$R$(1,2)\\
$Z_1$&$\Delta N$+1&$A_1$&$N_{1p\alpha}$&MeV&MEV&MeV&fm&fm\\\hline
 10 & 0    & 20   & 5      & 161  &    161   &  158   &  2.992(8)  & 2.92    \\
 11 & 1    & 23   & 5.5      & 187  &    187   &  185   &  2.94(6)   &   3.02  \\
 12 & 0    & 24   & 4      & 198  &    199   &  196   &  3.075(15) &   3.04  \\
 13 & 1    & 27   & 4.5      & 225  &    219   &  223   &  3.06(9)   &   3.13  \\
 14 & 0    & 28   & 4      & 237  &    238   &  234   &  3.14(4)   &   3.18  \\
 15 & 1    & 31   & 4.5      & 263  &    258   &  260   &  3.19(3)   &   3.26   \\
 16 & 0    & 32   & 4      & 272  &    277   &  271   &  3.240(11) &   3.31   \\
 17 & 1    & 35   & 4.5      & 298  &    297   &  297   &  3.388(17) &   3.39  \\
 18 & 0    & 36   & 5      & 307  &    306   &  307   &  3.327(15) &   3.43  \\
 19 & 1    & 39   & 5.5      & 334  &    332   &  333   &  3.408(27) &   3.53  \\
 18 & 4    & 40   & 2      & 344  &    343   &  349   &  3.393(15) &   3.38  \\
 19 & 5    & 43*  & 2.5      & 369  &    363   &  369   &            &   3.45  \\
 20 & 0    & 40   & 5      & 342  &    345   &  343   &  3.482(25) &   3.57  \\
 21 & 1    & 43*  & 5.5      & 367  &    371   &  369   &            &   3.63  \\
 20 & 2    & 42   & 3      & 362  &    370   &  364   &  3.505     &   3.52  \\
 21 & 3    & 45   & 3.5      & 388  &    394   &  390   &  3.550(5)  &   3.59   \\
 24 & 4    & 52   & 3      & 456  &    453   &  452   &  3.645(5)  &   3.73  \\
 25 & 5    & 55   & 3.5      & 482  &    477   &  477   &  3.680(11) &   3.79  \\
 28 & 2    & 58   & 5      & 506  &    502   &  502   &  3.760(10) &   3.96  \\
 29 & 3    & 61*  & 5.5      & 532  &    528   &  529   &            &   4.01  \\
 28 & 4    & 60   & 4      & 527  &    520   &  524   &  3.812(30) &   3.94  \\
 29 & 5    & 63   & 4.5      & 551  &    540   &  548   &  3.888(5)  &   3.99   \\
 30 & 4    & 64   & 4      & 559  &    560   &  557   &  3.918(11) &   4.02   \\
 31 & 5    & 67*  & 4.5      & 583  &    580   &  583   &            &   4.07  \\
 30 & 6    & 66   & 3      & 578  &    578   &  577   &  3.977(20) &   4.00   \\
 31 & 7    & 69   & 3.5      & 602  &    602   &  602   &            &   4.05\\
\\\hline
\end{tabular}}
\newpage
\begin{center}{Table 1. Continued.}\end{center} {\begin{tabular}{ccccccccc}\hline
$Z$&$\Delta N$&$A$&$N_{p\alpha}$&$E_{exp}$\cite{10}&$E$(4,5)&$E_{th}$(11)&$R_{exp}$\cite{11,12,13}&$R$(1,2)\\
$Z_1$&$\Delta N$+1&$A_1$&$N_{1p\alpha}$&MeV&MEV&MeV&fm&fm\\\hline
 40 & 10   & 90   & 3      & 784  &    788   &  778   &  4.28(2)   &   4.39  \\
 41 & 11   & 93   & 3.5    & 806  &    810   &  802   &  4.317(8)  &   4.43  \\
 50 & 18   & 118  & 3      & 1005 &     999  &  1002  &            &   4.72  \\
 51 & 19   & 121  & 3.5    & 1026 &    1022  &  1024  &  4.63(9)   &   4.76  \\
 50 & 20   & 120  & 2      & 1021 &    1026  &  1018  &  4.630(7)  &   4.71  \\
 51 & 21   & 123  & 2.5    & 1042 &    1046  &  1040  &            &   4.74   \\
 60 & 22   & 142  & 4      & 1185 &    1181  &  1177  &  4.993(35) &   5.02   \\
 61 & 23   & 145  & 4.5    & 1204 &    1201  &  1198  &            &   5.05  \\
 70 &  32  &  172 &  4     &  1393&     1390 &   1391 &            &    5.28 \\
 71 &  33  &  175 &  4.5   &  1412&     1410 &   1410 &   5.378(30)&    5.31 \\
 80 &  42  &  202 &  2+2   &  1595&     1580 &   1587 &   5.499(17)&    5.51 \\
 81 &  43  &  205 &  2+2.5 &  1616&     1615 &   1605 &   5.484(6) &    5.52 \\
 90 &  50  &  230 &  2+3   &  1755&     1752 &   1756 &            &    5.74 \\
 91 &  51  &  233*&  2+3.5   &  1772&     1776 &   1773 &            &    5.76 \\
 100&  52  &  252 &  3+4   &  1879&     1881 &   1881 &            &    5.95 \\
 101&  53  &  255*&  3+4.5   &  1896&     1901 &   1900 &            &    5.98 \\
 110&  60  &  281 &  2+3+3 &  2031&     2030 &   2037 &            &    6.15 \\
 111&  61  &  283*&  2+3+3.5 &      &     2047 &   2045 &            &    6.17  \\
 116&  72  &  304 &  2+3+3 &      &     2145 &   2147 &            &    6.26  \\
 117&  73  &  307(*?)&  2+3+3.5 &      &     2168 &   2157 &            &    6.28\\\hline
\end{tabular}}

\section{Conclusion}

The alpha - cluster  model \cite{3,4,5} has been developed to find some formulas for
calculation of the binding energies of $\beta$ - stable nuclei with using the notion
of core. This brings to a discovery that the specific density of the binding energy
of core for the nuclei of $\beta$-stable valley and those which are in its vicinity
can be a constant value equal to 2.57 $\rm MeV/fm^3$. The idea that the number of
excess neutrons is determined by the amount of $\alpha$-clusters of the core, which
has been approved before in terms of charge radii of nuclei \cite{3}, is approved
now in terms of the binding energies.

In the formulas (1) and (2) to calculate radii  the radius of an $\alpha$-cluster of
the core $r_\alpha = 1.60$ fm  is the only fitting parameter to describe the
experimental radii. So is the value $\rho = 2.57$ $\rm MeV/fm^3$ to describe the
experimental binding energies of all stable isotopes having core by the formulas (4)
and (5). Thus, it is clearly seen here that  the radius of one $\alpha$-cluster, as
well as the specific density of binding energy of core, are the constant values.
Then one has an opportunity to estimate the size of a nucleus from its experimental
binding energy.

In the heavy nuclei the growing Coulomb energy pushes out small compound clusters
from the core to the periphery of the nucleus. If for the stable nuclei the total
number of peripheral $\alpha$-clusters is within $2\div 5$, for the nuclei with $Z
\geq 80$ the number is within $5\div 8$.

The work is supported by international grant STCU  3081.

\end{document}